\documentclass[referee]{aa}
\usepackage{graphicx}

\def\gtrsim{\mathrel{\hbox{\rlap{\hbox{\lower4pt\hbox{$\sim$}}}\hbox{$>$}}}}
\def\ltrsim{\mathrel{\hbox{\rlap{\hbox{\lower4pt\hbox{$\sim$}}}\hbox{$<$}}}}
\def\eg {e.g. }
\def\etal {et al. }

\begin{document}

\title{New insights on the accuracy of photometric redshift
measurements}

\author{M. Massarotti\inst{1,}\inst{2}
\and A. Iovino\inst{2} 
\and A. Buzzoni\inst{2,}\inst{3}
\and D. Valls--Gabaud\inst{4}}

\offprints{M. Massarotti.\\
\email{massarot@brera.mi.astro.it;\\
iovino@brera.mi.astro.it; buzzoni@tng.iac.es; dvg@ast.obs-mip.fr}}

\institute{Osservatorio Astronomico di Capodimonte, Via Moiariello 16, 80131 Napoli, Italy
\and
Osservatorio Astronomico di Brera, Via Brera 28, 20121 Milano, Italy
\and
Telescopio Nazionale Galileo, P.O. Box 565, 38700 Santa Cruz de La Palma (TF), Spain
\and
CNRS UMR 5572, Observatoire Midi-Pyr\'en\'ees, 14, Ave. E. Belin,
 31400 Toulouse, France  
}

\date{Received ; Accepted}

\abstract{We use the deepest and most complete redshift catalog
currently available (the Hubble Deep Field (HDF) North supplemented by
new HDF South redshift data) to minimize residuals between photometric
and spectroscopic redshift estimates.  The good agreement at
$z_\mathrm{spec} < 1.5$ shows that model libraries provide a good
description of the galaxy population.  At $z_\mathrm{spec} \geq 2.0$,
the systematic shift between photometric and spectroscopic redshifts
decreases when the modeling of the absorption by the interstellar and
intergalactic media is refined. As a result, in the entire redshift
range $z \in [0, 6]$, residuals between photometric and spectroscopic
redshifts are roughly halved.  For objects fainter than the
spectroscopic limit, the main source of uncertainty in photometric
redshifts is related to photometric errors, and can be assessed with
Monte Carlo simulations.   }

\authorrunning{M. Massarotti et al.}
\titlerunning{Accuracy of photometric redshifts} 
\maketitle 

\section{Introduction}\label{sec_1}

Reliable spectroscopy of galaxies fainter than $I_\mathrm{AB} \sim 24$
is still an unattainable goal even with new-generation telescopes.  To
estimate the redshift for a majority of these objects one must
therefore rely on multicolor observations in order to reconstruct, at
very low resolution, the apparent spectral energy distribution (SED).

Much efforts have been devoted, in the recent literature, to improve
the accuracy of the photometric approach compared with direct
spectroscopy.  Photometric redshift techniques have been extensively
applied, in particular, to the ultra--deep images of the Hubble Deep
Field North (HDFN, Williams \etal1996, see also Ferguson, Dickinson
and Williams 2000) to study the luminosity function, the evolution of
the star formation rate (SFR) density, and the clustering properties
of the galaxies observed there (Madau \etal1996; Connolly \etal1997;
Sawicki \etal1997; Franceschini \etal1998; Pascarelle \etal1998;
Connolly \etal1998; Miralles \& Pell\'o 1998; Roukema \etal1999,
Arnouts \etal1999).

The sample of relatively bright HDFN objects with reliable
spectroscopic redshifts is steadily increasing and now amounts to
about 150 objects, a value that nearly doubles the initial list of
Sawicki \etal(1997).  About 20\% of these objects are beyond $z \geq
2.0$, thus allowing a more robust statistical comparison between
spectroscopic and photometric estimates of $z$ even for the most
distant objects in the field.

We previously discussed (Massarotti \etal2001a, Paper I hereafter) the
performance of the SED fitting method as a function of the template
set adopted to reproduce galaxy colors.  Our results suggested that,
as long as both normal and starburst galaxies are taken into account
in the reference templates, the photometric redshift at $z < 1.5$ is
constrained by a reliable detection of the 4000 \AA\ Balmer break in
the target SED. At larger redshifts, on the contrary, the main
constraint is set by modeling the effects of the interstellar (ISM)
and intergalactic (IGM) absorption, which modulate the SEDs at
different wavelengths.  In this paper, we further discuss some
important aspects of this problem, and suggest an improved model for
the ISM and IGM absorption which sensibly increases the accuracy of
the method at $z \gtrsim 2$.

Photometric redshifts are estimated by comparing the observed
broad-band colors of galaxies with the library of model fluxes
provided by the code of Buzzoni (1989, 1995, 2001; hereafter
BUZ). Results obtained with the codes of Bruzual \& Charlot (1993, BC)
and Fioc \& Rocca--Volmerange (1997, FRV) are also reported, since
these theoretical models have also been extensively used in the
literature.

The key steps of our procedure are summarized in Sect.~2, where we
also introduce the model libraries.  The redshift database used in our
analysis is presented in Sect.~3, comparing the $z_\mathrm{spec}$
distribution with our $z_\mathrm{phot}$ output. The ISM and IGM
effects are more specifically dealt with in Sect.~4, while in Sect.~5
we discuss the match between models and observations even at the
faintest magnitude limit of the HDFN data. Section 6 summarizes our
conclusions.

\section{Input ingredients and fitting algorithm}\label{sec_2_1}

As in Paper I, the photometric catalog of Fern\'andez--Soto
\etal(1999) will be used as a reference in our analysis. It is made up
by 1067 objects identified from the HST WFPC2 images in the four
photometric bands, namely $U_\mathrm{300}$, $B_\mathrm{450}$,
$V_\mathrm{606}$, and $I_\mathrm{814}$, and completed with infrared
observations in the Johnson $J, H$, and $K$ bands with the IRIM camera
at KPNO (Dickinson 1998).

Galaxy model libraries have been assembled following the strategy of
Paper I. A notable change in the present calculations involves a
slight enlargement of the set of starburst models in order to cover a
wider range of physical parameters and ensure a more accurate
description of the colors of the most distant galaxies.

To the Leitherer \etal(1999) starburst sequence of Paper I, namely for
$t = 50$, 100, 500, and 800 Myr, with $Z_\odot$ and Salpeter IMF, we
added here two supplementary models at $t = 10$ and 250 Myr, and
included evolution for $Z_\odot/20$ metallicity in addition to the
solar case previously considered.

In order to account for recent star formation superimposed on
quiescent red (elliptical) galaxies, we also considered a new set of
templates. These are obtained adding the 50 Myr starburst spectrum
(with Salpeter IMF and $Z_\odot$) to simple stellar population models
for a multi-component fit to the galaxy SED. In this way the overall
color distribution of early type galaxies at $z < 1.5$, as found by
Franceschini \etal(1998) and Rodighiero \etal(2001), is much better
accounted for.

As a first step, the ISM and IGM absorptions have been modeled in as
in Paper I.  Galaxy model fluxes have been corrected for ISM effects
following Calzetti (1999).  As in Paper I, the $E(B-V)$ color excess
has been set as a free fitting parameter which could take the discrete
values 0.0, 0.05, 0.1, 0.2, 0.3, 0.4.  The IGM absorption as a
function of redshift follows the analaysis by Madau (1995).

According to our SED fitting method, we searched for the ``best
template'' solution as which minimizes the general $\chi^2$ function
(see Eq.~(1) of Paper I). In the multi-component fit to the galaxy SED
the $\chi^2$ functional takes the form: 

\begin{equation}
\label{form_1}
\chi^2=\sum_{i=1}^{N}{{[f^\mathrm{obs}_i-s_\mathrm{ssp}\
f^\mathrm{temp}_\mathrm{i,ssp}(z)-s_\mathrm{burst}\
f^\mathrm{temp}_\mathrm{i,burst}(z)]^2}\over{\sigma_i^2}}\,,
\end{equation}

\noindent
where $f^\mathrm{temp}_\mathrm{i,ssp}$ and
$f^\mathrm{temp}_\mathrm{i,burst}$ are the template fluxes of the
quiescent galaxy model and of the burst model, respectively, scaled by
$s_\mathrm{ssp}$ and $s_\mathrm{burst}$, and $N$ is the number of
filters.

For galaxies with $z_\mathrm{spec}$ available ($N_\mathrm{g}$) we
define two simple statistical moments:

\begin{equation}
\label{form_2}
\overline {\Delta z} ={\sum^{N_\mathrm{g}}
{{(z_\mathrm{spec}-z_\mathrm{phot})}\over{N_\mathrm{g}}}}\,,
\end{equation}
and
\begin{equation}
\label{form_3}
\sigma_\mathrm{z}^2 = \sum^{N_\mathrm{g}}
{{[(z_\mathrm{spec}-z_\mathrm{phot})-\overline {\Delta z}]^2}\over{N_\mathrm{g} -1}}\,,
\end{equation}
to study residuals between photometric and spectroscopic redshift
values.

Catastrophic outliers (that is, galaxies with
$|z_\mathrm{spec}-z_\mathrm{phot}|>1.0$) are clipped from the galaxy
distribution to compute representative values for $\overline{\Delta
z}$ and $\sigma_\mathrm{z}^2$.

\section{The spectroscopic sample}\label{sec_3}

Cohen \etal(2000) provided an exhaustive compilation of all
spectroscopic redshifts measured with the Keck telescope in the HDFN
proper and flanking fields.  The work includes previously published
data, new redshifts measurements, and corrections to erroneous values
given in the literature. This list represents the deepest and most
complete redshift catalog for the HDFN to date.

Fern\'andez--Soto \etal(2001) matched photometric and redshift
catalogs in the HDFN proper. They cross--correlated coordinates and
magnitudes, resulting in a list of 146 sources (see Table~2 therein).
We adopt here their naming convention and follow their considerations
regarding some galaxies with discordant $z_\mathrm{spec}$ vs.\
$z_\mathrm{phot}$ values, except for the following few objects.

{\bf HDF36441--1410}: we assume that the spectroscopic measure
$z_\mathrm{spec}=2.267$ is correct. Fern\'andez--Soto \etal(2001)
claimed that features identifications in the spectrum are compatible
with their estimated $z_\mathrm{phot}=0.01$, but {\bf HDF36441--1410}
is a $U$ drop--out star--forming galaxy
($U_\mathrm{300}-B_\mathrm{450} \sim 1.7$,
$V_\mathrm{606}-I_\mathrm{814} \sim 0.0$). These colors (blue in
$V_\mathrm{606}-I_\mathrm{814}$, red in
$U_\mathrm{300}-B_\mathrm{450}$) can be explained only as consequence
of the IGM absorption of UV photons in the $U_\mathrm{300}$ band, and
are incompatible with $z \sim 0$ starburst colors.

{\bf HDF36446--1227}: we prefer the spectroscopic value
$z_\mathrm{spec}=2.035$ proposed by Lowenthal \etal(1997), to the
(revised) value $z_\mathrm{spec}=2.500$ reported by Cohen \etal(2000)
and accepted by Fern\'andez--Soto \etal(2001). In the redshift range
$z_\mathrm{spec} \in [2, 2.3]$ there is no galaxy bluer than {\bf
HDF36446--1227} in $U_\mathrm{300}-B_\mathrm{450}$, and according to
the starburst templates of our models its optical colors are
incompatible with $z_\mathrm{spec}>2.3$ (see Sect.~\ref{ISM}).

{\bf HDF36396--1230}, {\bf 36494--1317}, and {\bf 36561--1330}: for
these three galaxies we follow Cohen (2001), who adopts the redshift
values given by Fern\'andez--Soto \etal(2001).

\begin{figure*}
\resizebox{\hsize}{!}{\includegraphics{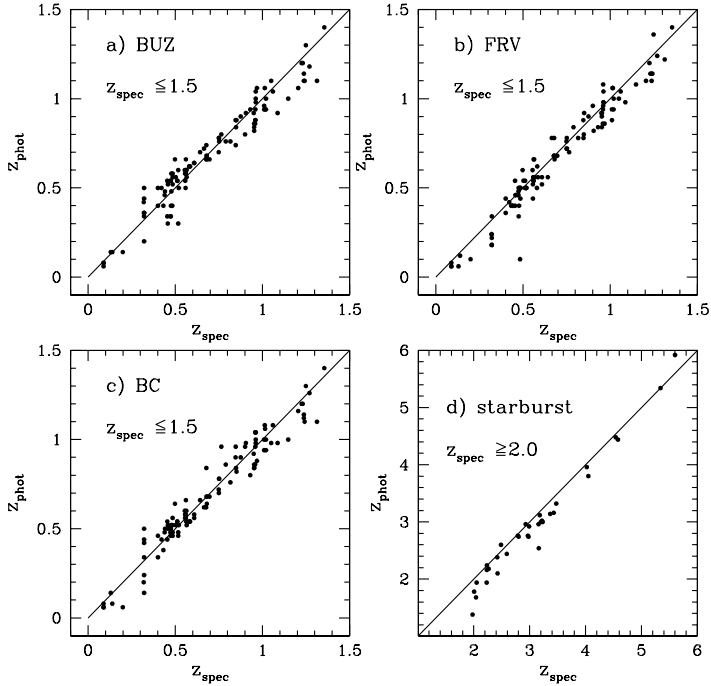}}
\caption{Comparison between spectroscopic and photometric redshifts
according to different template libraries (see the text for
acronyms). The solid line is for $\Delta z=0.0$. Note the different
scale for panel {\it d}.}
\label{fig-1}
\end{figure*}

{\bf HDF36478--1256}: with $z_\mathrm{spec}=2.931$, this galaxy was
tagged as a catastrophic outlier in all previous works (see \eg
Fern\'andez--Soto \etal1999; Bolzonella \etal2000; Fern\'andez--Soto
\etal2001; Fontana \etal2000).  Arnouts \etal(1999) suggest that the
photometry could be incorrect because of the complex source
morphology.  Given the unclear situation and the large discrepancy for
this galaxy (our photometric redshift estimate is
$z_\mathrm{phot}=0.26$) we excluded it from the catalog.

Six new galaxies from the HDF--South (HDFS) have also been added to
our list. Three of them are from Glazebrook \etal(2001) with
$z_\mathrm{spec} = 0.580$, 0.582, and 0.696, and three are from
Cristiani \etal(2000) with $z_\mathrm{spec}=0.565$, 2.79, and 3.2
respectively.  The photometry was obtained from the web page of the
Stony Brook group (see {\sf http://www.ess.sunysb.edu/astro/hdfs}) in
the four WFPC2 filters, and in the $J, H$, and $K$ infrared bands by
the SOFI camera at NTT (Da Costa \etal2001).

Our final spectroscopic catalog contains 146 objects, 34 of them with
$z_\mathrm{spec} \geq 2$.

\subsection{Comparison between $z_\mathrm{spec}$ and $z_\mathrm{phot}$}\label{sec_4}

In Fig.~\ref{fig-1} we compare spectroscopic and photometric redshifts
obtained with different sets of reference galaxy models.

At $z_\mathrm{spec}<1.5$ (panels {\it a,b,c}) the agreement is
excellent, regardless of the template set chosen. Dispersion values of
the photometric redshift estimates are $\sigma_\mathrm{z} = 0.078$,
0.074, 0.072 for the BUZ, FRV and BC libraries, respectively, while
systematic errors are $\overline{\Delta z}=0.007$, 0.026, 0.004. There
are no galaxies with $|z_\mathrm{spec}-z_\mathrm{phot}|>0.5$.

Since in this redshift range the accuracy of the method depends mainly
on the ability of models to reproduce the 4000 \AA\ Balmer break along
the whole Hubble morphological sequence (including starburst
galaxies), one can conclude that all model libraries contain enough
information to match properly the SEDs.

For the same sample, at $z_\mathrm{spec}<1.5$, Fontana and
collaborators obtained $\sigma_\mathrm{z} = 0.084$ and
$\overline{\Delta z}=0.033$ (see the web page {\sf
http://www.mporzio.astro.it/HIGHZ/HDF.html}).  Fern\'andez--Soto
\etal(1999) obtained $\sigma_\mathrm{z} = 0.116$ and $\overline{\Delta
z}=-0.007$, further improved to $\sigma_\mathrm{z} = 0.107$ and
$\overline{\Delta z}=-0.004$ after including starburst galaxies in
their template library (Fern\'andez--Soto \etal2001). Bolzonella
\etal(2000) obtained $\sigma_\mathrm{z} \sim 0.082$, using a smaller
spectroscopic sample, that did not account for the new spectroscopic
redshifts from Cohen and collaborators.

The situation becomes more complex at $z_\mathrm{spec} \geq 2$ (see\
panel {\it d} of Fig.~\ref{fig-1}). The dispersion increases to
$\sigma_\mathrm{z} = 0.177$ and there are two objects with
$|z_\mathrm{spec}-z_\mathrm{phot}|>0.5$.  In this redshift range
almost all galaxies are best fit by starburst models, and we therefore
focussed on the Leitherer \etal(1999) templates.

Part of the larger scatter can certainly result from the fainter
magnitudes (lower signal--to--noise ratio), but a systematic shift
($\overline{\Delta z}=0.156$) is significant in panel {\it d}, and
cannot be explained in terms of photometric errors only.  The tendency
to underestimate redshifts for $z_\mathrm{spec} \geq 2$ seems to be a
common feature of all photometric studies found in the literature,
regardless of the specific choice of theoretical and/or empirical
template library.

The only exception in this regard are the results of Fern\'andez--Soto
\etal(2001) which show systematics in the opposite sense, with
$\overline{\Delta z}=-0.136$.  These authors did {\it not} however
take into account explicitely the ISM reddening modulation of
the template SEDs (the average amount of dust that the empirical
templates do contain may not be entirely representative of the dust
content of galaxies at high redshifts). As we noted in Paper I, this
causes a significant overestimate (by about 10--20\%) of $z$ at large
distances, because the IGM absorption, which increases with redshift,
is forced to mimic color reddening.

The presence of dust in starburst galaxies at high redshift has been
confirmed by analyzing galaxy colors with the SED fitting method (see
Massarotti \etal2001b, and references therein). Meurer \etal(1999)
adopted an entirely empirical approach based on the correlation
between color reddening, dust absorption and far infrared flux for
local starburst galaxies (supposed to be also valid at high
redshifts). According to their results (see also Meurer \etal1997;
Sawicki \& Yee 1998), at $<z>=2.75$ a color--luminosity correlation is
already in place: more luminous galaxies (i.e.  galaxies with
available spectroscopic redshifts) have a larger dust content.

The observed systematic shift in the $z_\mathrm{spec}$
vs. $z_\mathrm{phot}$ distribution of Fig.~\ref{fig-1}, calls
therefore for a revised analysis of the ISM and IGM effects in the $z
\gtrsim 2$ data.

\section{A revised approach to IGM and ISM absorption}\label{new}

In Paper I we showed that at high redshifts the SED fitting method
provides stable estimates of galaxy redshifts, even when using an
extremely poor template library, while the critical ingredients are
dust reddening and IGM attenuation. It is important to explore in some
detail the physical processes of photon absorption in both cases.

\subsection{The IGM opacity as a function of redshift}\label{IGM}

Following Madau (1995), the Ly$\alpha$ forest contribution to the
attenuation of the SED caused by the IGM absorption can be written in
terms of an effective optical depth as 

\begin{equation}
\label{form_4}
\frac{F_\mathrm{obs}(\lambda,z)}{F_\mathrm{intr}(\lambda)}=
e^{-\tau_\mathrm{eff}}\,,
\end{equation}

\noindent
with

\begin{equation}
\label{form_5}
{{\tau_\mathrm{eff}} = {A \times (\lambda_\mathrm{obs}/\lambda_\mathrm{\alpha})^{1+\gamma}}}\,,
\end{equation}

\noindent
where $F_\mathrm{obs}(\lambda,z)$ and $F_\mathrm{intr}(\lambda)$ are
the observed and intrinsic fluxes respectively,
$\lambda_\mathrm{obs}=\lambda \times (1+z)$, and
$\lambda_\mathrm{\alpha}=1216$ \AA.

Press \etal(1993), in their analysis of the distribution of Ly$\alpha$
forest lines in a sample of 29 quasars at $z \in [3.1,4.5]$, found
$(A,~\gamma) = (0.0036,~2.46)$, which are the values usually adopted
in the literature.

In a recent paper, Scott \etal(2000) studied a sample containing 97
quasars at $z \in [1.7,3.6]$, a redshift range complementary to that
studied by Press \etal(1993). Considering absorption lines above the
threshold $W>0.16$~\AA, they obtain $(A,~\gamma) = (0.00759,~1.35)$.
The IGM transmission at $z<3$ is therefore larger than the one
expected by extrapolating the Press \etal(1993) results at these
redshifts.

We decided to adopt the Scott \etal(2000) results to describe the IGM
absorption at $z<3$ and the Press \etal(1993) results at $z>4$.  In
the intermediate redshift range (i.e. between $3 < z < 4$), the $A$
and $\gamma$ coefficients of Eq.~(5) have been interpolated linearly
imposing as boundary limits $(A,~\gamma) = (0.00759,~1.35)$ at $z = 3$
and $(A,~\gamma) = (0.0036,~2.46)$ at $z = 4$.

A useful way to match our adopted theoretical scheme with the
observations is to derive from Eq.~(5) the expected IGM flux
attenuation ($D_\mathrm{A}$) in the spectral range $\lambda \in
[1050,1170]$ \AA\ as a function of redshift. This can be directly
compared with the observed deficit in quasars fluxes due to the
accumulated absorption of the Ly$\alpha$ lines in the IGM.  By
definition,

\begin{equation}
\label{form_6}
D_\mathrm{A}(z)=1-\frac{\int_{1050}^{1170}F_\mathrm{obs}(\lambda,z)d\lambda}
{\int_{1050}^{1170}F_\mathrm{intr}(\lambda)d\lambda}\,.  
\end{equation}

\begin{figure}
\resizebox{\hsize}{!}{\includegraphics{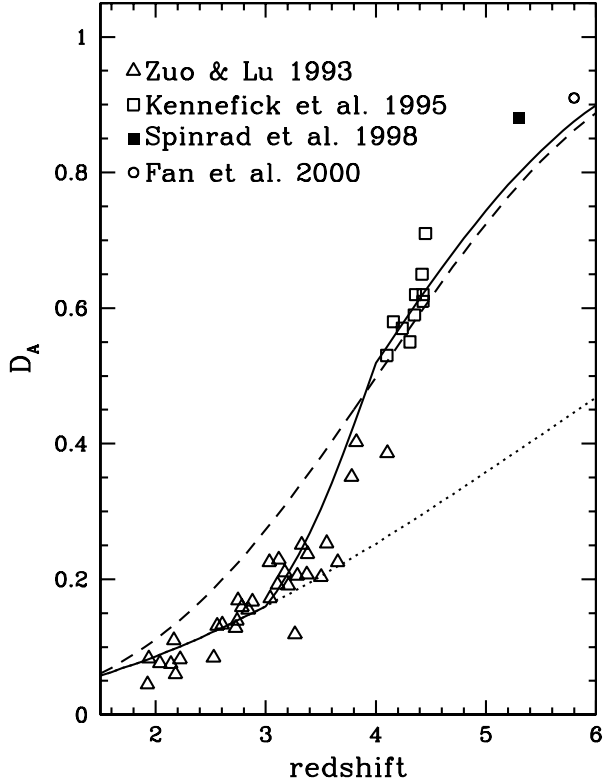}}
\caption{The deficit $D_\mathrm{A}$ in quasars fluxes due to the
accumulated absorption of the Ly$\alpha$ forest.  The dashed line is
obtained by describing the IGM opacity according to the prescriptions
by Madau (1995), while the dotted line are those by Scott
\etal(2000). The solid line is the model adopted in this paper.
References for the data points are indicated. The small difference
between the Madau values of $D_\mathrm{A}$ and ours at $z > 4$ is due
to the contribution of Ly limit systems (Madau 1995, $Eq.~(13)$),
which was not included in Madau's formulation.}
\label{fig-2}
\end{figure}

\noindent
Our results are summarized in Fig.~\ref{fig-2}, which collects
observations from different sources. The data cannot be described by a
single power law, since the slope of $D_\mathrm{A}(z)$ changes in the
interval $z \in [3,4]$. The Press \etal(1993) solution (adopted by
Madau 1995) sensibly overestimates the IGM opacity at $z<3.5$; the
opposite happens with the Scott \etal(2000) results when extending
them beyond $z>3.5$. Note, on the contrary, that our simple
parameterization fits nicely the $D_\mathrm{A}(z)$ evolution over the
entire range of redshift, from $z = 2$ to $z = 6$.

\begin{figure}
\resizebox{\hsize}{!}{\includegraphics{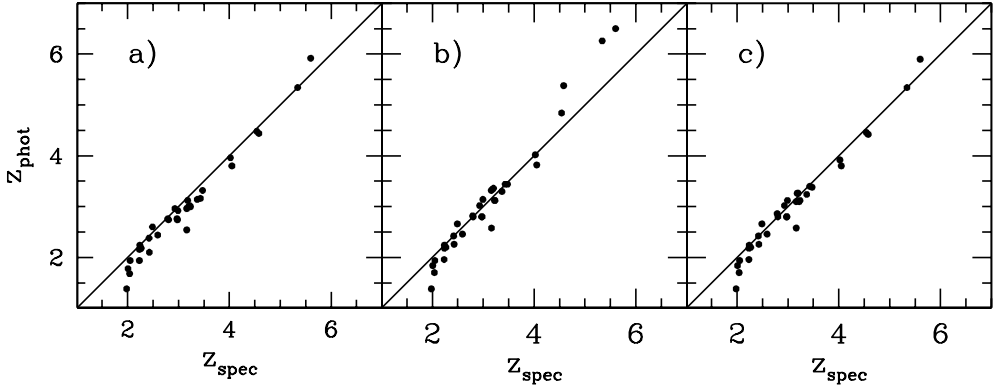}}
\caption{Comparison between spectroscopic and photometric redshifts at
$z_\mathrm{spec} \geq 2$. Panel {\it a}: the IGM opacity follows the
prescription by Madau (1995); panel {\it b}: by Scott \etal(2000);
panel {\it c}: the formulation discussed in this paper. The solid line
is for $\Delta z=0.0$. It should be noted that when the Scott
\etal(2000) results are adopted, the IGM opacity is underestimated at
$z > 3.5$, and, as a consequence, photometric redshifts overestimate
the spectroscopic ones.}
\label{fig-3}
\end{figure}

We repeated the photometric redshift estimate for the $z_\mathrm{spec}
\geq 2.0$ objects adopting our new prescriptions for the IGM
absorption.  For $z > 2.5$, only one outlier (namely {\bf
HDF36512--1349} at $z_\mathrm{spec}=3.162$) is found at
$|z_\mathrm{spec}-z_\mathrm{phot}|>0.5$. In the Cohen \etal(2000) list
the spectroscopic redshift of this object is reported as (very)
uncertain.

Excluding this source, the dispersion decreases to
$\sigma_\mathrm{z}=0.129$ (cf.~Fig.~\ref{fig-3} {\it c}). More
importantly, the tendency to underestimate galaxy redshift almost
disappears ($\overline{\Delta z}=0.042$ at $z_\mathrm{spec} \geq
2.5$). This is not true however in the range $z \in [2, 2.5]$, where
the results are still problematic.

\subsection{The ISM opacity and the bump at 2175 \AA}\label{ISM}

In the redshift range $z_\mathrm{spec} \in [2, 2.5]$, the 912 \AA\
Lyman break has not yet entered the $U_\mathrm{300}$ band and
therefore galaxies can still be detected in this filter, although with
a major dimming effect by the IGM absorption.

The $U_\mathrm{300}$ magnitude does not effectively constrain $z$ in
the SED fitting method, because of its poor signal-to-noise ratio
\footnote{Note that, paradoxically, the same
spectral signature which defines so effectively the $U$ drop--out
sample at $z \geq 2$ causes the $U_\mathrm{300}$ information to be rather useless in
the SED fitting method.}.  The $\chi^2$ merit function is
mainly modulated in this redshift range by the optical information coming from
the $B_\mathrm{450}$, $V_\mathrm{606}$ and $I_\mathrm{814}$ bands.  It is therefore
plausible that any bias in the $z$ estimate is linked in some way with the
$B_\mathrm{450}-V_\mathrm{606}$ and $V_\mathrm{606}-I_\mathrm{814}$ colors.

\begin{figure}
\resizebox{\hsize}{!}{\includegraphics{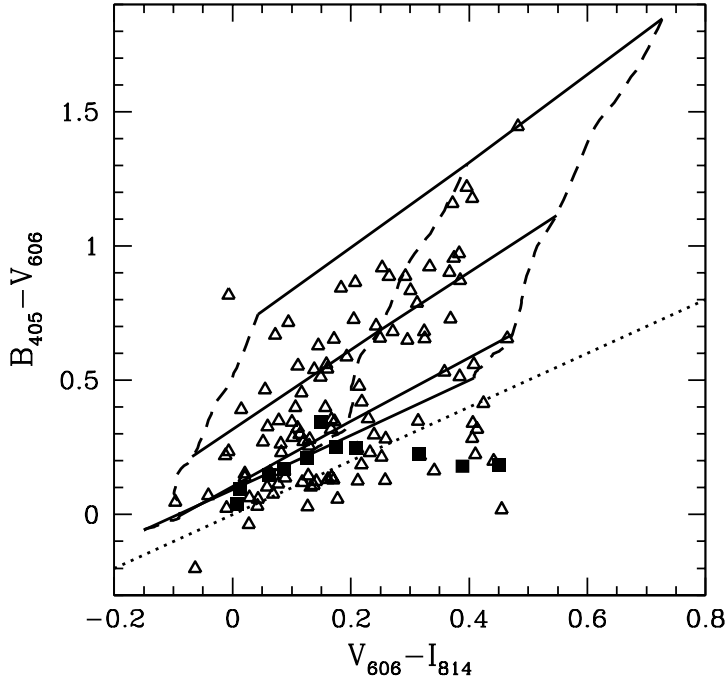}}
\caption{Observed colors of $U$ drop--out galaxies (triangles),
galaxies with spectroscopic redshifts in the range $z \in [2,2.5]$
(filled squares), and model colors of starburst galaxies in the
redshift range $z \in [2,3.5]$ (see the text for details). Dashed
lines show the color evolution of templates for increasing redshift
(from bottom to top) at $E(B-V)=0.0,0.2,0.4$ (from left to
right). Solid lines define {\it reddening vectors} at
$z=2.0,2.5,3.0,3.5$ (from bottom to top), that is, the loci of
templates at fixed redshift with variable amounts of dust. The dotted
line corresponds to
$B_\mathrm{450}-V_\mathrm{606}=V_\mathrm{606}-I_\mathrm{814}$.}
\label{fig-4}
\end{figure}

At $z \sim 2$ the $B_\mathrm{450}$ band is probing the rest-frame
emission at about $1500 \pm 300$~\AA, while the $V_\mathrm{606}$ band
samples the wavelength range $2000 \pm 300$~\AA, and the
$I_\mathrm{814}$ band that at $2700 \pm 300$~\AA.  Between 1500 and
2700 \AA, the Leitherer \etal(1999) starburst models indicate that the
galaxy spectral slope steepens with increasing wavelength so that, in
general, $(B_\mathrm{450}-V_\mathrm{606}) >
(V_\mathrm{606}-I_\mathrm{814})$ in the AB mag scale.

This is confirmed by Fig.~\ref{fig-4}, where the color diagram for the
100 Myr starburst model with Salpeter IMF and $Z_\odot$ is shown.
Dust reddening can only strengthen this effect: dust absorption
reddens $B_\mathrm{450}-$~$V_\mathrm{606}$ more than
$V_\mathrm{606}-I_\mathrm{814}$ (see in Fig.~\ref{fig-4} the {\it
reddening vectors}). Moreover, at $z>2.5$ the Lyman series enters the
$B_\mathrm{450}$ band and the $B_\mathrm{450}-V_\mathrm{606}$ color is
further reddened by the IGM absorption.

In Fig.~\ref{fig-4} we also plotted the HDFN $U$ drop--out sample
(triangles) selected according to the prescriptions of Meurer
\etal(1999) ($U_\mathrm{300}-B_\mathrm{450} \geq 1.3$,
$V_\mathrm{606}-I_\mathrm{814}<0.5$, and $B_\mathrm{450} \leq 26.8$).
A large fraction (24\%) of these galaxies displays observed colors
$(B_\mathrm{450}-V_\mathrm{606}) < (V_\mathrm{606}-I_\mathrm{814})$
and 3 out of the 11 sources at $z_\mathrm{spec} \in [2,2.5]$ are among
these.  Note that for galaxies with spectroscopic reddshifts, the
color errors are $\leq 0.02 $, so it is difficult to explain such
large deviations from models in terms of photometric errors only.  To
account in a consistent way for the deviating
$B_\mathrm{450}-V_\mathrm{606}$ and $V_\mathrm{606}-I_\mathrm{814}$
colors, one could think that a dimming in the $V_\mathrm{606}$ flux is
intervening in the data, at about the rest-frame wavelength
corresponding to the 2175 \AA\ bump of the dust extinction law in the
Milky Way (Seaton 1979) or the Large Magellanic Cloud (Fitzpatrick
1986).

\begin{figure}
\resizebox{\hsize}{!}{\includegraphics{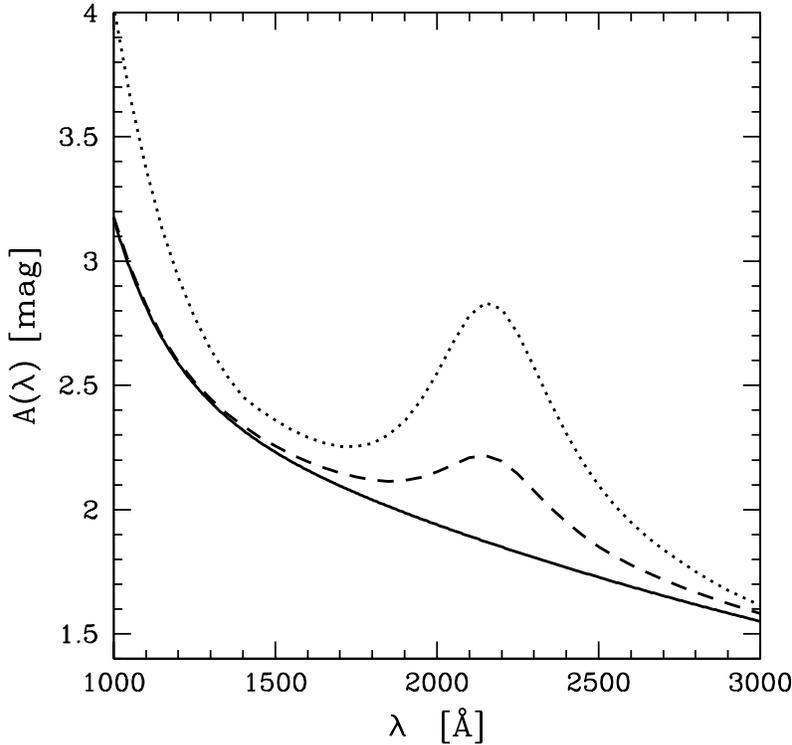}}
\caption{Differential dust absorption (for $E(B-V)=0.2$) according to
Calzetti's law (solid line), a superposition of Calzetti's law with a
2175 \AA\ bump of amplitude $\alpha=0.25$ (dashed line), and the
Seaton (1979) extinction curve for the Milky Way (dotted line).}
\label{fig-5}
\end{figure}

We tried to superimpose the bump $C(\lambda)$ on Calzetti's law, in
the form of a Drude profile according to Fitzpatrick \& Massa (1988,
see also Calzetti \etal1994):

\begin{equation}
\label{form_7}
C(\lambda)=\alpha \times \tau_\mathrm{B}^\mathrm{l} {{x^2} \over
{(x^2-x^2_\mathrm{0})^2+ \gamma x^2}}\,,
\end{equation}

\noindent
where $x=\lambda^{-1}$ $(\mu m^{-1})$,
$x_\mathrm{0}=\lambda^{-1}_\mathrm{0}=4.6$ $\mu m^{-1}$ is the central
wavelength of the dust bump, $\gamma=0.95$ $\mu m^{-1}$ is its FWHM,
and $\tau_\mathrm{B}^\mathrm{l}=E(B-V)/(0.935 \times 0.44)$ is the
Balmer optical depth, as inferred from the difference in the optical
depth between the nebular emission lines H$\alpha$ and H$\beta$ (see
Calzetti \etal1994; Calzetti 2000).  Introducing the bump, the
differential dust attenuation curve takes the form
$A(\lambda)=[k(\lambda)+C(\lambda)] \times E(B-V)$, where $k(\lambda)$
follows Calzetti's law, and $C(\lambda)$ comes from Eq.~(7).

In Fig.~\ref{fig-5}, the differential dust attenuation curve is shown
(for $E(B-V)=0.2$) without the bump (solid line) and with a bump of
intrinsic depth $\alpha=0.25$ (dashed line). For comparison the Milky
Way extinction curve is also shown (Seaton 1979), where the intrinsic
depth of the bump is $\alpha \sim 0.8$ ($\alpha \sim 0.4$ for the
Large Magellanic Cloud, see for example Calzetti \etal1994, Fig. 13).

\begin{figure}
\resizebox{\hsize}{!}{\includegraphics{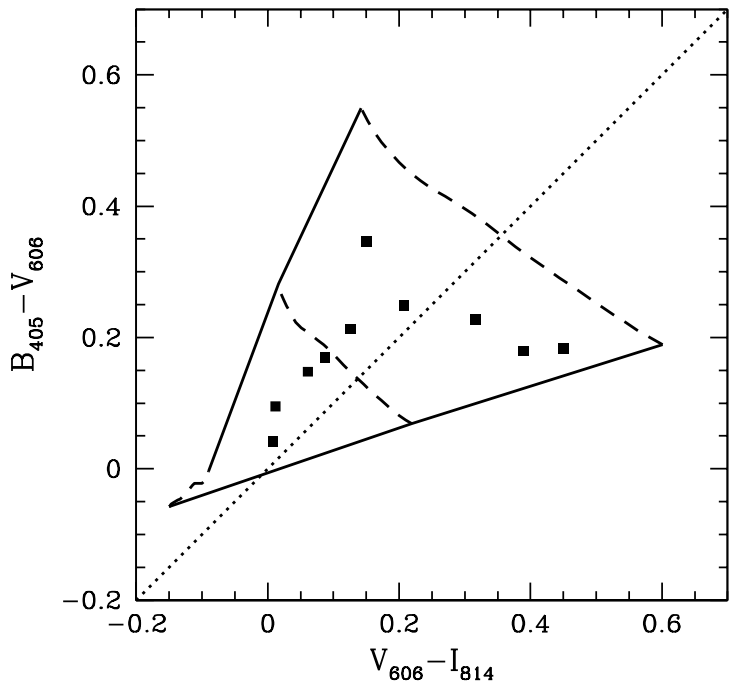}}
\caption{Observed colors of spectroscopic galaxies (squares) and model
colors of starburst galaxies in the redshift range $z \in
[2,2.5]$. Dashed lines and solid lines have the same meaning as in
Fig.~\ref{fig-4}, but in this case reddening in colors due to dust
absorption is obtained superimposing on Calzetti's law a 2175 \AA\
bump with amplitude $\alpha=0.25$. Again, the dotted line corresponds
to $B_\mathrm{450}-V_\mathrm{606}=V_\mathrm{606}-I_\mathrm{814}$. As
consequence of the rotation of the {\it reddening vectors}, model
colors are able to describe galaxy colors with
$(B_\mathrm{450}-V_\mathrm{606}) < (V_\mathrm{606}-I_\mathrm{814})$.}
\label{fig-6}
\end{figure}

When the 2175 \AA\ bump is superimposed on the dust attenuation law,
the locus occupied by starburst galaxies in the
$(B_\mathrm{450}-V_\mathrm{606})$ vs.\
$(V_\mathrm{606}-I_\mathrm{814})$ plane (in the range $z \in [2,2.5]$)
broadens to include galaxy colors with
$B_\mathrm{450}-$~$V_\mathrm{606}$~$<V_\mathrm{606}-I_\mathrm{814}$
(see Fig.~\ref{fig-6}).  Adding the 2175 \AA\ bump ($\alpha=0.1,
0.25$) enables us to better reproduce galaxy colors in the range
$z_\mathrm{spec} \in [2,2.5]$ and significantly improves the agreement
between photometric and spectroscopic redshift estimates. While
$\sigma_\mathrm{z} = 0.206$, $\overline{\Delta z}=0.167$ (without the
bump), the values decrease when the bump is allowed to
$\sigma_\mathrm{z} = 0.094$, $\overline{\Delta z}=0.060$.  For
the three galaxies (see Fig. 6) which adopt a 2175 \AA\ bump value
different from zero in their fit, the $\chi^2$ value is, on the mean,
4 times smaller than the one obtained with no bump, confirming the
statistical significance of our improvement.

The question of the presence of the 2175 \AA\ bump in the dust
attenuation law is still hotly debated in the literature.  Gordon
\etal(1997) claimed that radiative transfer effects associated with
the spatial distribution of dust cannot (alone) decrease the feature
below observational detectability.  Therefore, the lack of a 2175 \AA\
bump in a starburst galaxy must be explained as a consequence of
physical and chemical properties of dust grain. Reversing the
argument, in a starburst with LMC--like dust, geometrical effects
alone would certainly not prevent its possible detection.

In contrast, Granato \etal(2000) found, with their dust model for
starbursts, that the 2175 \AA\ bump is absent due to purely
geometrical effects.  In the far-UV, the global attenuation in
starburst models is dominated by the component associated with
molecular clouds, where stars are born, and from which they gradually
escape. Clouds have such large optical depths that the ultraviolet
light from the stars inside is completely absorbed, and the shape of
the attenuation curve does not contain information on the UV-optical
properties of grains.
 
On the other hand, relying on the same theoretical framework, the
results of Silva \etal(1998) lead us to conclude that virtually all
the stars embedded in dark molecular clouds escape the regions of
harder absorption in less than $10^8$yr (cf. also Buzzoni 2001 on this
subject). In such cases, the cirrus component dominates and a (weak)
2175 \AA\ bump should eventually appear in the dust attenuation law.

Calzetti \etal(1994) did explicitly avoid the 2175 \AA\ bump
wavelength region when calibrating their starburst attenuation law, as
the possible presence of absorption bump would destroy a linear
correlation between the UV power law slope $\beta$ and
$\tau_\mathrm{B}^\mathrm{l}$. When fitting {\it a posteriori} the
residuals between observed spectra and the appropriate linear law
($\log F(\lambda)=\beta \log \lambda + {\rm const}$) according to the
Fitzpatrick \& Massa (1988) parameterization, they obtained a mean
value $\alpha =0.07$ for the 2175 \AA\ bump amplitude. It should be
noted, though, that among the objects in their sample with
$\tau_\mathrm{B}^\mathrm{l} \geq 0.25$ ($E(B-V) \geq 0.1$, see again
Calzetti \etal1994, Fig. 13) -- that is, those where the presence of
the feature is easier to detect-- about $\sim 35 \%$ are compatible
with $\alpha =0.25$.

Our results support the presence of the 2175 \AA\ bump in some
galaxies observed in the spectroscopic sample: values of $\alpha$ up
to 0.25 allow a better description of galaxy colors and
redshifts. Higher-quality spectra for these controversial objects are
needed to assess beyond any reasonable doubt the actual presence of
the bump.

\begin{figure}
\resizebox{\hsize}{!}{\includegraphics{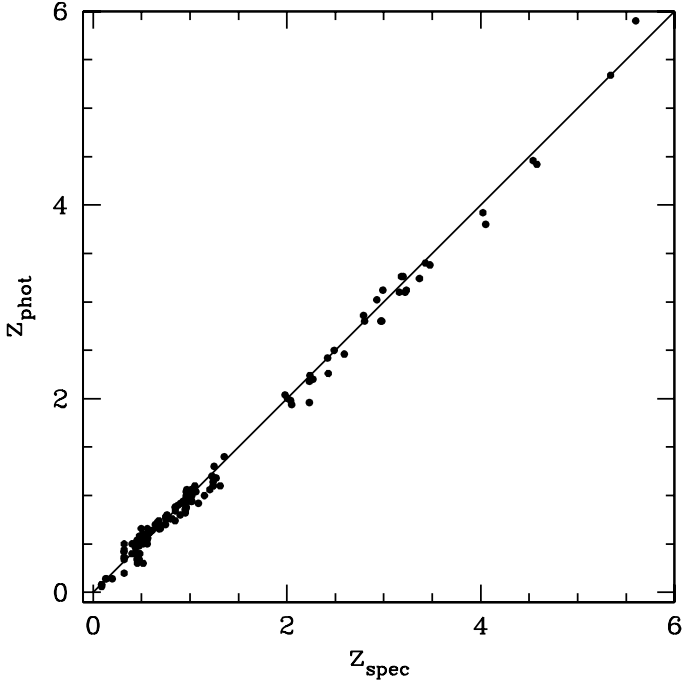}}
\caption{Comparison between spectroscopic redshifts and photometric
estimates obtained with the BUZ library plus Leitherer \etal(1999)
models in the entire redshift interval $z \in [0, 6]$. At
$z_\mathrm{spec} \geq 2$, the ISM and IGM opacities follow the rules
introduced in this work. The solid line is for $\Delta z=0.0$.}
\label{fig-7}
\end{figure}

\section{Updating our results}

The inclusion of the refinements discussed above in the SED model
libraries significantly improved the agreement between photometric and
spectroscopic redshifts at $z_\mathrm{spec} \geq 2.0$. The dispersion
decreases to $\sigma_\mathrm{z}=0.117$, while the systematic error is
$\overline{\Delta z}=0.048$ (see Fig.~\ref{fig-7}), with no
catastrophic outliers.

Using the same sample in the redshift range $z_\mathrm{spec} \geq
2.0$, Fontana and collaborators obtained $\sigma_\mathrm{z} = 0.232$,
$\overline{\Delta z}=0.144$ (with one catastrophic outlier),
Fern\'andez--Soto \etal(1999) obtained $\sigma_\mathrm{z} = 0.299$,
$\overline{\Delta z}=0.186$ (with two catastrophic outliers), while
for Fern\'andez--Soto \etal(2001) $\sigma_\mathrm{z} = 0.234$ and
$\overline{\Delta z}=-0.136$ (with one catastrophic outlier). With a
smaller spectroscopic sample, Bolzonella \etal(2000) estimated
$\sigma_\mathrm{z} \sim 0.36$ (with one catastrophic outlier).

A further small improvement can be obtained by taking into account
dust opacity in the Galaxy along the line of sight to the
HDFN. Introducing the effects of dust absorption on model fluxes in
the observed frame following the Milky Way extinction curve (Seaton
1979) with $E(B-V)=0.012$, as inferred from the maps of Schlegel
\etal(1998), we obtain minor changes in the redshift estimates, though
in the direction of improving the agreement with spectroscopic
redshifts.  Using the BC code, for example, we find $\sigma_\mathrm{z}
= 0.071$, $\overline{\Delta z}=0.002$ at $z_\mathrm{spec} < 1.5$, and
$\sigma_\mathrm{z} = 0.117$, $\overline{\Delta z}=0.036$ at
$z_\mathrm{spec} \geq 2.0$.

\begin{figure}
\resizebox{\hsize}{!}{\includegraphics{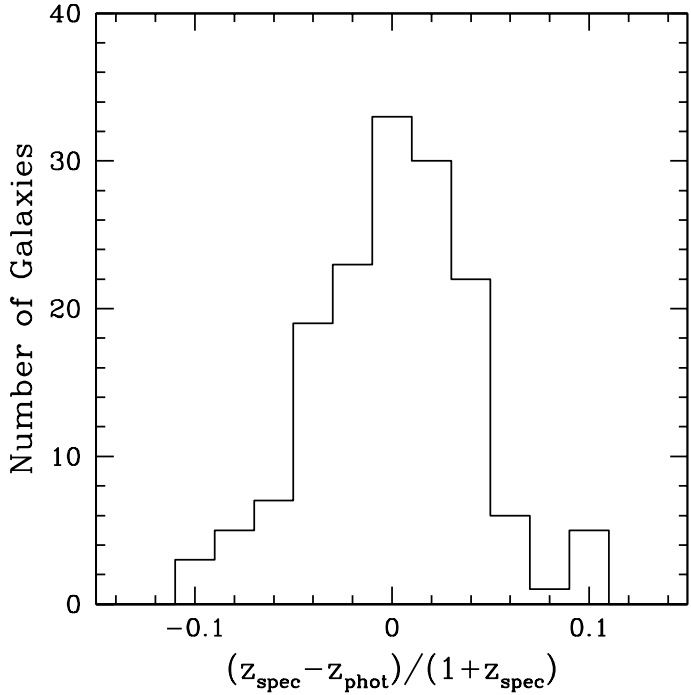}}
\caption{The distribution of residuals between $z_\mathrm{spec}$ and
$z_\mathrm{phot}$, normalized to $(1+z_\mathrm{spec})$.}
\label{fig-8}
\end{figure}

Dawson \etal(2001) recently published 12 new redshifts within the
HDFN. Adding this subset to our sample (but excluding galaxies 2-600.0
and 4-236.0, which have a very tentative redshift), we obtain
$\sigma_\mathrm{z}/(1+z_\mathrm{spec}) \sim 0.040$ on the largest deep
spectroscopic sample available up to date (155 objects).  This is a
factor of $2.0-2.5$ better than previous estimates in the
literature. We find $\sigma_\mathrm{z} = 0.070$, $\overline{\Delta
z}=0.001$ at $z_\mathrm{spec} < 1.5$, and $\sigma_\mathrm{z} = 0.116$,
$\overline{\Delta z}=0.034$ at $z_\mathrm{spec} \geq 2.0$.
Figure~\ref{fig-8} shows the distribution of residuals between
$z_\mathrm{spec}$ and $z_\mathrm{phot}$, normalized to
$(1+z_\mathrm{spec})$. It suggests that all estimated redshifts are
confined in the range $-0.1 \leq
(z_\mathrm{spec}-z_\mathrm{phot})/(1+z_\mathrm{spec}) \leq 0.1$,
confirming the reliability of the SED fitting procedure.

\subsection{The faint limit}\label{sec_6}

Our results state the accuracy of the SED fitting method when
applied to the bright spectroscopic sample. When galaxies are fainter
and photometric errors increase the accuracy obviously decreases.

We used a simple Monte Carlo method to separate the effects of
observational uncertainties from the effects due to model
incompleteness (``template mismatch'') in defining the distribution of
residuals between photometric and spectroscopic redshifts. We
generated 100 copies of the spectroscopic sample, adding to each
galaxy flux a correction randomly extracted from the Gaussian
distribution of its photometric error.

We compared, for each object, the redshift obtained from the measured
photometry $z_\mathrm{phot}$ and that from the bootstrapped galaxy
photometry $z_\mathrm{sim}$. Due to photometric errors, the dispersion
value is $\sigma_\mathrm{z}^\mathrm{err}/(1+z_\mathrm{phot}) \sim
0.031$, and the systematic error goes to zero.

To quantify the contribution of model incompleteness (and only that)
to the dispersion, we assume that the two sources of error are
uncorrelated. In this case,

\begin{equation}
\label{form_8}
\sigma_\mathrm{z}^2 \sim
(\sigma_\mathrm{z}^\mathrm{err})^2+(\sigma_\mathrm{z}^\mathrm{temp})^2\,,
\end{equation}

\noindent
and we obtain $\sigma_\mathrm{z}^\mathrm{temp}/(1+z_\mathrm{spec})
\sim 0.025$.

Since the residuals introduced by template incompleteness are
negligible compared to the photometric redshift instability caused by
observational uncertainties in the intermediate and faint galaxy
sample (see Paper I, Table 2), we can conclude that, at the faint
limit, Monte Carlo methods are sufficient to estimate errors in the
SED technique.

\begin{figure}
\resizebox{\hsize}{!}{\includegraphics{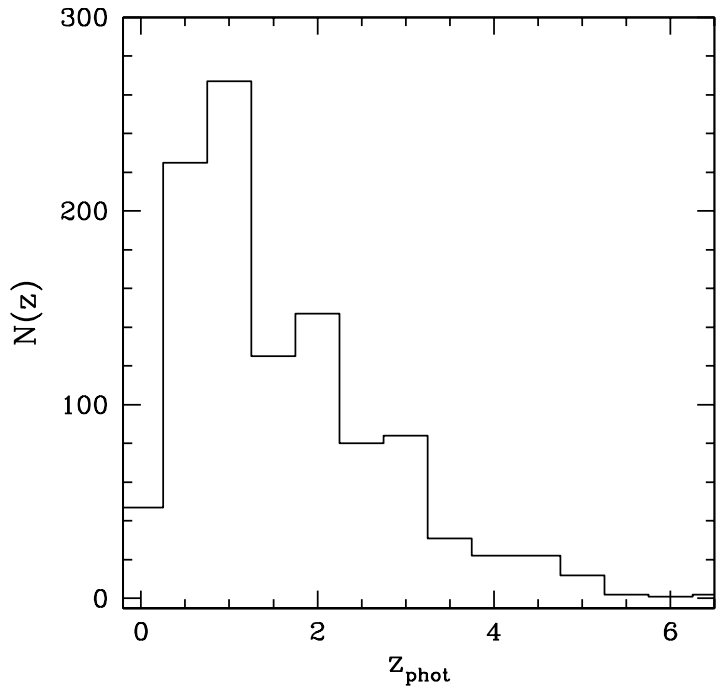}}
\caption{The redshift distribution of HDFN galaxies}
\label{fig-9}
\end{figure}

In Fig.~\ref{fig-9} we show the redshift distribution of HDFN galaxies
obtained by applying our photometric redshift code to the entire
catalog of Fern\'andez--Soto \etal(1999). The well known peak at $z
\sim 0.9$ is clearly visible and there is a gradual decline at larger
redshifts, with a tail at very high redshift $ z > 4.5$.

\section{Discussion and conclusions}\label{ris}

We have explored the accuracy of the SED fitting method to estimate
photometric redshifts by comparing its results to spectroscopic
measures from the deepest and most complete redshift catalog to date
in the HDFN (Cohen \etal2000) complemented by new HDF South redshift
data.  We used as photometric dataset the Fern\'andez--Soto
\etal(1999) catalog for the HDF North, and for the HDF South the
catalogue from the Stony Brook's group (web page {\sf
http://www.ess.sunysb.edu/astro/hdfs}).  The libraries of galaxy
colors were supplied by the theoretical codes of Buzzoni (2001),
Bruzual \& Charlot (1993), and Fioc \& Rocca--Volmerange (1997) to
reproduce normal galaxy SEDs for different morphological types, and by
the theoretical code by Leitherer \etal (1999) to describe starburst
SEDs.

A very good agreement between spectroscopic and photometric redshifts
is found at $z_\mathrm{spec}<1.5$, independently of the template set
used. Model libraries include enough information to properly reproduce
galaxy SED as a function of redshift from the ultraviolet to
near--infrared. At $z_\mathrm{spec} \geq 2$ the usual recipes for
quantifying ISM and IGM absorption yield photometric redshift
estimates that are systematically smaller than the spectroscopic ones
($\overline{\Delta z}=0.156$), while the dispersion increases to
$\sigma_\mathrm{z}=0.177$, confirming previous studies.

We focussed our attention on the prescriptions used to describe photon
absorption in the IGM and ISM, since these processes have been shown
(Paper I) to be the deciding factors in the redshift estimate at
$z_\mathrm{spec} \geq 2$.

In the redshift range $z \in [1.5, 3]$ the Madau (1995) prescription
(adopting the Press \etal1993 values) overestimates IGM opacities with
respect to the values obtained by Scott \etal(2000) on a larger quasar
sample.  The observed deficit in quasar fluxes due to the cumulative
absorption by the Ly$\alpha$ lines of the IGM cannot be described with
a single power law because the slope of $D_\mathrm{A}(z)$ changes in
the interval $z \in [3,4]$.  We adopted the Scott \etal(2000) values
at $z<3$ and Press \etal(1993) values at $z>4$, and a linear
interpolation between these two estimates in the range $z \in
[3,4]$. In this way at $z>2.5$ the tendency to underestimate galaxy
redshifts disappears and at the same time the dispersion decreases
($\sigma_\mathrm{z}=0.129$).

In the range $z_\mathrm{spec} \in [2, 2.5]$, 3 objects in the
spectroscopic sample show optical colors
($B_\mathrm{450}-V_\mathrm{606} < V_\mathrm{606}-I_\mathrm{814}$)
incompatible with intrinsic model colors for starburst galaxies and
with the dust reddening absorption law of Calzetti (1999). The common
feature is a flux dimming in the $V_\mathrm{606}$ band, which samples
the restframe wavelength region $\lambda \sim 2000 \pm 300$ \AA\ at $z
\sim 2$.  Adding to the dust attenuation law a weak ($\alpha=0.1,
0.25$) 2175 \AA\ bump, enables template colors to reproduce the
observed colors of high redshift starburst galaxies in the
$(B_\mathrm{450}-V_\mathrm{606})$ vs.\
$(V_\mathrm{606}-I_\mathrm{814})$ plane.

When these improvements are implemented, the dispersion between
photometric and spectroscopic redshifts at $z_\mathrm{spec} \geq 2.0$
becomes $\sigma_\mathrm{z}=0.116$, and the systematic error
$\overline{\Delta z}=0.034$. In the entire redshift interval $z \in
[0, 6]$, we obtain the value $\sigma_\mathrm{z}/(1+z_\mathrm{spec})
\sim 0.04$. This is an improvement by a factor $\sim 2.0-2.5$ over
previous estimates in the literature.

\section{Acknowledgments}\label{sec-3}
We would like to thank the referee A. Fern\'andez-Soto for his careful
reading of the manuscript and useful comments.  This work received
partial financial support by Fondazione Cariplo and COFIN grant
00-02-016. A.I. and D.V.-G. thank ESO for financial support during a
period spent at ESO-Santiago, where part of this work was done.

\end{document}